\documentclass[lettersize, journal]{IEEEtran}
\usepackage{amsmath,amsfonts}
\usepackage{algorithmic}
\usepackage{algorithm}
\usepackage{array}
\usepackage[caption=false,font=normalsize,labelfont=sf,textfont=sf]{subfig}
\usepackage{textcomp}
\usepackage{stfloats}
\usepackage{url}
\usepackage{verbatim}
\usepackage{graphicx}
\usepackage{cite}
\usepackage{multirow}
\usepackage{orcidlink}
\usepackage{xcolor} 
\usepackage{graphicx} 
\usepackage{flushend}

\newcommand{\HUMP}{\emph{HuMP-CAT}}

\begin{document}

\title{Leveraging Cross-Attention Transformer and Multi-Feature Fusion for Cross-Linguistic Speech Emotion Recognition
}

\author{Ruoyu Zhao\orcidlink{0009-0008-9254-1918}, Xiantao Jiang\orcidlink{0000-0002-2226-482X}, F. Richard Yu \orcidlink{0000-0003-1006-7594}, Fellow, IEEE, 
 
Victor C.M. Leung \orcidlink{0000-0003-3529-2640}, Fellow, IEEE,  Tao Wang\orcidlink{0000-0002-0865-0062}, and Shaohu Zhang\orcidlink{0000-0001-8985-515X}
~\IEEEmembership{}

\thanks{Ruoyu Zhao and Xiantao Jiang are with the Department of Information Engineering, Shanghai Maritime University, NO.1550, Haigang Ave., Shanghai 201306, China (email: xtjiang@shmtu.edu.cn).

F. Richard Yu is with the Department of Systems and Computer Engineering, Carleton University, Ottawa, ON KIS 5B6, Canada (e-mail:richard.yu@carleton.ca).

V. C. M. Leung is with the Department of Electrical and Computer Engineering, The University of British Columbia, Vancouver, BC V6T 1Z4, Canada (e-mail:vleung@ece.ubc.ca).

Tao Wang is with Stanford School of Medicine, Stanford University, 450 Jane Stanford Way
Stanford, California 94305, USA (e-mail:taowang9@stanford.edu).

Shaohu Zhang is with the Department of Mathematics and Computer Science, The University of North Carolina at Pembroke. 1 University Drive Pembroke, North Carolina 28372, USA (e-mail:shaohu.zhang@uncp.edu)

Xiaotao Jiang and Shaohu Zhang are the corresponding authors.
}

}


\maketitle

\begin{abstract}
Speech Emotion Recognition (SER) plays a crucial role in enhancing human-computer interaction. Cross-Linguistic SER (CLSER) has been a challenging research problem due to
significant variability in linguistic and acoustic features of different languages. In this study, we propose a novel approach \HUMP, which combines HuBERT, MFCC, and prosodic characteristics. These features are fused using a cross-attention transformer (CAT) mechanism during feature extraction. Transfer learning is applied to gain from a source emotional speech dataset to the target corpus for emotion recognition. We use IEMOCAP as the source dataset to train the source model and evaluate the proposed method on seven datasets in five languages (e.g., English, German, Spanish, Italian, and Chinese).  We show that, by fine-tuning the source model with a small portion of speech from the target datasets, \HUMP~achieves an average accuracy of 78.75\% across the seven datasets, with notable performance of 88.69\% on EMODB (German language) and 79.48\% on EMOVO (Italian language). Our extensive evaluation demonstrates that \HUMP~outperforms existing methods across multiple target languages.

\end{abstract}


\section{Introduction}
\IEEEPARstart{S}{peech} emotion recognition (SER) has attracted considerable attention for its potential applications in various areas, including human-computer interaction, healthcare, education, entertainment, transportation systems, and the Internet of Things (IoT) \cite{el2011survey,rana2019automated,wani2021comprehensive}. SER enable IoT devices or applications to better understand the user's emotions and to respond more appropriately to the user's psychological state. For example, in smart assistants (such as Siri, Alexa, etc.), if anger or happiness in the user's voice can be recognized, the system can adjust the voice tone and respond appropriately to enhance the interactive experience. 

SER involves on identifying and understanding emotional states through vocal cues such as pitch, prosody, and Mel frequency cepstral coefficients (MFCC) \cite{wang2022systematic}. Traditional SER systems that leverage classical machine learning, such as support vector machine (SVM) \cite{lin2005speech}, naive Bayes classifier \cite{wang2015speech}, and \(k\)-nearest neighbor (KNN) \cite{lanjewar2015implementation}, have been limited by language-specific constraints and have struggled to generalize emotional patterns across linguistic speech. The emerging field of cross-linguistic speech emotion recognition (CLSER) seeks to transcend these limitations by developing robust, adaptive models that can interpret emotional states irrespective of linguistic variations.

 However, CLSER faces several challenges. The first key challenge is the scarcity of large and balanced cross-linguistic emotion datasets. Most existing datasets are heavily biased towards a few languages such as English and European languages, which create significant representational gaps. The second challenge is to identify effective features for emotion recognition. Although numerous features such as pitch, prosody, MFCC, linear prediction cepstral coefficients (LPCC), and Gammatone frequency cepstral coefficients (GFCC)\cite{wani2021comprehensive}, have been shown to correlate with emotions, it remains to be explored how to combine these features optimally to improve CLSER performance. The third challenge involves improving the generalization and versatility of recognition across languages. Due to significant variations in language such as accent, culture, and other factors between training and testing datasets, the majority of SER systems are limited to a single language corpus but do not generalize well to other language corpus.

Recent advances in deep learning and representation learning have provided opportunities for SER due to their robust feature representation capability, capacity to manage complex features, ability to learn from unlabeled data, and scalability with larger datasets. Various deep learning models, such as convolutional neural networks (CNN), deep neural networks (DNN), long- and short-term memory networks (LSTM) and transformers, have been applied for automatic SER \cite{khalil2019speech,9,10}. 

Self-supervised speech representation learning (SSRL) is a form of unsupervised learning that seeks to capture rich representations from the input speech signal itself \cite{14,15,16}. which have been widely adopted in various speech processing tasks, including SER \cite{tao2022self,zhang2023voicepm, ma2024emobox}. HuBERT~\cite{18} and Wav2vec~\cite{17} are prominent examples of SSRL methods applied in SER. These approaches have also been successfully applied in multilingual and cross-corpus SER \cite{19,20}. HuBERT integrates representation information from speech recognition models to train a multimodal emotion recognition framework \cite{21}. Sharma et al. \cite{22} fine-tuned Wav2vec2 for multilingual and multitask learning. Although these methods utilize the capacity of SSRL models and transformer layers to extract suitable representations for SER, they do not explore the specific contribution of these transformer layers to CLSER. Moreover, the role of these layers in the fusion with other acoustic features of CLSER remains unexplored.

In this paper, we propose \textbf{\emph{HuMP-CAT}}, a novel CLSER framework that integrates \textbf{\underline{Hu}}BERT, \textbf{\underline{M}}FCC, \textbf{\underline{P}}rosody, and \textbf{\underline{CAT}}. \HUMP~leverages multi-feature fusion and cross-attention transformer (CAT) models to create a more universal approach to understanding emotions in diverse linguistic speech. The main contributions of this paper are as follows.
\begin{itemize}
     \item We propose a novel CLSER framework that integrates CAT models and multi-feature fusion to achieve higher SER accuracy compared to existing methods.
 
    \item \HUMP~leverages transfer learning by using only 10\% or 20\% of the speaker data to fine-tune the source model, achieving an impressive average accuracy of 78.75\% across seven diverse datasets. This demonstrates the model's capacity to generalize SER with small training data.

    \item We perform a comprehensive evaluation across five languages (e.g., English, German,
Spanish, Italian, and Chinese) in seven datasets, examining the performance of various SSRL models and the combination of speech characteristics. Our extensive experiments demonstrate \HUMP's generalization and versatility in CLSER.
\end{itemize}

The remainder of this paper is organized as follows. Section \ref{sec:related_work} provides a brief review of related work on SER and CLSER. Section \ref{sec:architecture} presents the proposed multi-feature fusion method based on the CAT mechanism and discusses each module of the approach.  Section \ref{sec:experiments} describes the dataset, experimental procedure, and results. Conclusions are provided in Section \ref{sec:concusion}.

\section{RELATED WORK}
\label{sec:related_work}
\begin{table}[]
\caption{Summary of major abbreviations\label{tab:table1}}
\centering
\begin{tabular}{ccc}
\hline
Abbreviation & Description and full name                         &  \\ \hline
SER          & Speech Emotion Recognition                        &  \\
CLSER        & Cross-Linguistic Speech  Emotion Recognition      &  \\
MFCC         & Mel-Frequency Cepstral Coefficient              &  \\
HuBERT       & Hidden-Unit BERT                                  &  \\
SSRL         & Self-Supervised Speech Representation Learning    &  \\
CAT          & Cross-Attention Transformer                       &  \\
AM-Softmax   & Additive Margin Softmax                           &  \\
LSTM         & Long Short-Term Memory Networks                 &  \\
DSCNN        & Deep Stride Convolutional Neural Networks       &  \\
UA           & Unweighted Accuracy                               &  \\
WA           & Weighted Accuracy                                 &  \\ \hline
\end{tabular}
\end{table}
\begin{figure*}[!t]
\centering
\includegraphics[width=0.75\linewidth]{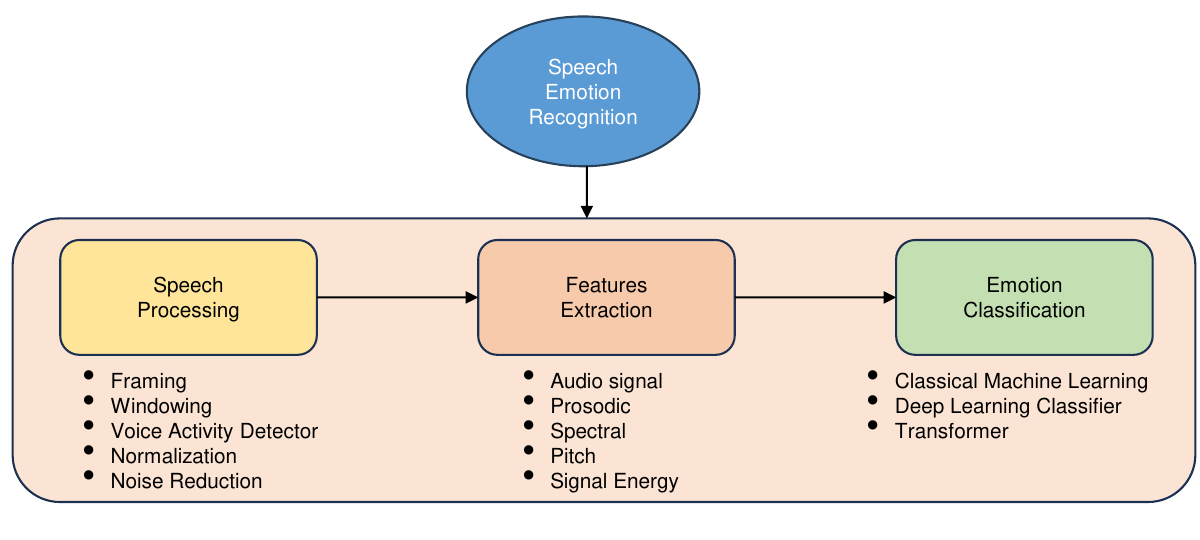}
\caption{General process of SER.}
\label{fig_1}
\end{figure*}
In this paper, we present a novel CLSER framework that integrates
CAT models and multi-feature fusion. This section reviews the related work on various aspects of existing SER and CLSER approaches. We also highlight the differences and contributions of our proposed framework compared to the existing methods. Table \ref{tab:table1} lists the abbreviations and their corresponding meanings that often appear in the paper.
\subsection{Speech Emotion Recognition}

As shown in Fig.\ref{fig_1}, the computational pipeline of SER comprises three critical modules: speech preprocessing, feature extraction, and sentiment classification, each playing a pivotal role in transforming raw acoustic signals into meaningful emotional representations. The preprocessing stage is the prerequisite for the subsequent extraction of features, including speech framing, windowing, normalization, and noise reduction. Feature extraction represents the transformation stage, where preprocessed signals are converted into discriminative emotional representations. This module extracts both time- and frequency-domain features, including prosodic, spectral, pitch, and energy-based features, and passes the extracted feature to the emotion classification module for SER. 

Traditional machine learning methods \cite{lin2005speech,wang2015speech,lanjewar2015implementation}, such as SVM, Random Forest (RF), and Hidden Markov Models (HMM), depend heavily on manually selected features, such as MFCCs, prosodic features (e.g., pitch and energy contours), and statistical models, to classify emotions in speech \cite{2022systematicsurvey}. Although effective to some extent, these approaches are limited by the quality and relevance of hand-crafted features, which may fail to capture the complex, high-dimensional nature of speech signals. 

In contrast, deep learning (DL) techniques \cite{khalil2019speech,9,10,24}, including CNNs, Recurrent Neural Networks (RNNs), LSTMs, and Autoencoders, can automatically learn hierarchical feature representations directly from raw or minimally processed speech data. This ability to extract and model intricate temporal and spatial patterns has improved significantly. For example, Wu et al. \cite{24} introduced sequential capsule networks that excel in capturing spatial and contextual information, further enhancing the performance of SER. 

SSRL is an unsupervised learning designed to extract rich and meaningful representations directly from raw speech signals \cite{14,15,16,tao2022self}. Using the structure inherent in speech data, SSRL methods eliminate the need for extensive labeled data sets and have been adopted in SER \cite{tao2022self}. Popular SSRL frameworks, such as HuBERT \cite{18} and Wav2vec \cite{17}, have demonstrated significant potential in SER, offering robust pre-trained representations that enhance downstream model performance, even in resource-constrained emotion dataset. For instance, Zhang et al.\cite{25} used an encoder based on Transformer, which can be pretrained by using a large amount of unlabeled audio from various datasets and is able to learn more general and robust acoustic representations. Li et al. \cite{26} used contrastive predictive coding to learn important representations in unlabeled datasets for SER. 

\subsection{Cross-Linguistic SER}
CLSER has been a challenge problem in the field of speech processing. Early approaches \cite{eybencross2010,schuller2011using, zehra2021cross} relied on prosodic features such as pitch, intensity, and rhythm, which were combined with classical machine learning models such as SVM and RF. While these methods showed initial promise, they struggled with generalization across languages due to significant variability in linguistic and acoustic features. For instance, Zehra et al. \cite{zehra2021cross} trained their model using an Urdu dataset \cite{latif2018crossUrdu} and tested it on datasets in other languages, including English, German, and Italian, achieving accuracies ranging from 50\% $\sim$ 60\%. With advances in deep learning, Braunschweiler et al. \cite{braunschweiler2021study} applied the CNN-RNN-ATT model, achieving an improved accuracy of 60\% to 70\%.  

Recent studies have also explored the use of self-supervised speech representation learning (SSRL) such as HuBERT \cite{18} and WavLM \cite{chen2022wavlm}, which have shown better performance over earlier models such as Wav2Vec2\cite{17}. However, few studies have combined these representations with Cross-Linguistic strategies. Pepino et al. \cite{pepino2021emotion} evaluated the performance of Wav2vec2 for SER systems, but did not use a Cross-Linguistic strategy. Unlike traditional spectral and cepstral representations, SSRL models can learn richer and more robust feature embeddings directly from raw audio data, offering promising improvements for SER tasks. Therefore, emerging directions in CLSER include integrating SSRL techniques with domain adaptation or transfer learning methods to bridge the gap between source and target languages. Combining SSRL models with advanced architectures, such as transformer-based systems, may further enhance the performance of CLSER by more effectively capturing acoustic and linguistic variability.

\subsection{Distinction with Related Work}
Unlike previous studies, our proposed framework \HUMP~combines the recent SSRL HuBERT model and CAT with two additional sets of speech characteristics including MFCC and prosodic features. This multi-feature fusion enhances the performance of SSRL in CLSER by leveraging complementary information from both traditional and learned representations.
In addition,  to mitigate complex databases that negatively impact performance, we use IEMOCAP, a large emotion English dataset, as the training corpus. This avoids data scarcity issues while ensuring a robust and representative model pretraining phase.

To extend the applicability of our system to CLSER, we employ transfer learning to adapt trained models, leveraging both the feature fusion module and the classifier to low-resource target languages. Fine-tuning is then performed on a small subset of the target corpus, allowing the system to learn language-specific characteristics while preserving the generalization capabilities gained during pretraining.

%
\section{Architecture Design of \HUMP}
\label{sec:architecture}
\subsection{Overview}
 \HUMP~is comprised three main components including Feature Extraction, Multi-feature Fusion and Classification. As shown in Fig.\ref{fig:overview}, \HUMP~extracts the audio feature using the HuBERT model, MFCC, and prosody. The HuBERT model computes a 768-dimensional representation of the audio and passes the speech representation to the convolutional block with a kernel size of 10*18 and a stride of 4*3.  CATs are used to fuse the extracted features from the three input blocks and reduce the dimensionality of the input features.
\subsection{HuBERT}
Hidden-Unit BERT (HuBERT) is a self-supervised speech representation learning framework \cite{35}, drawing inspiration from the BERT model \cite{36} and Wav2vec2. Its architecture combines a CNN encoder with 12 transformer blocks. HuBERT employs the K-means algorithm to perform an offline clustering operation on 39-dimensional MFCC features, along with their first-order (\(\Delta\)) and second-order differences ( \(\Delta\Delta\)). The framework calculates the prediction loss using a combination of masked masked $Loss_m$ and unmasked $Loss_u$ time steps, aligned with the Wav2vec2 approach. The loss of cross-entropy for the masked, unmasked and overall time steps is defined as follows:
\cite{35, chakhtouna2024unveiling}.
\begin{equation}
\label{eq7}
Loss_{m}(f,S,S^\prime,Z)=\Sigma_{t\in{\cal S^\prime}}\log p_{f}(z_{t}|\tilde{S},t)
\end{equation}

\begin{equation}
\label{eq8}
Loss_{u}(f,S,S^\prime,Z)=\Sigma_{t\not\in{\cal S^\prime}}\log p_{f}(z_{t}|\tilde{S},t)
\end{equation}
\begin{equation}
\label{eq9}
Loss=\lambda Loss_{m}+(1-\lambda)Loss_{u}
\end{equation}
where, $S^\prime$ denotes the set of indices in the input sequence $S$ that need to be masked, while $Z$  represents the target sequence derived through K-means clustering. The masked embedding sequence is denoted by \(\tilde{S}\) if \( t\in S^\prime\), otherwise \( t\notin S^\prime\). 
 The overall loss function  \( Loss\)  is a weighted sum of the masked \( Loss_m\)  and unmasked \( Loss_u\)  losses by using a weighted constant \(\lambda\)
 
HuBERT model outputs 768-dimensional audio representations. During training, 
random masking is applied to contiguous time steps of the local encoder's representations
 Labels for pre-training are initially generated by clustering MFCC features using K-means, while later iterations leverage latent features derived from the preceding iterations to refine the target quality.
 \HUMP~leverages the pre-train model trained on 960 hours of the Librispeech dataset and selects the transformer blocks that are most robust for SER as features. Specifically, the 1st and 9th transformer layers representing acoustics and phonetic information are selected \cite{38}. \HUMP~combines these features through attention-based fusion to derive the final HuBERT feature representation.

\subsection{Cross-Attention Transformer}
\begin{figure}[!t]
\centering
\includegraphics[width=0.95\linewidth]{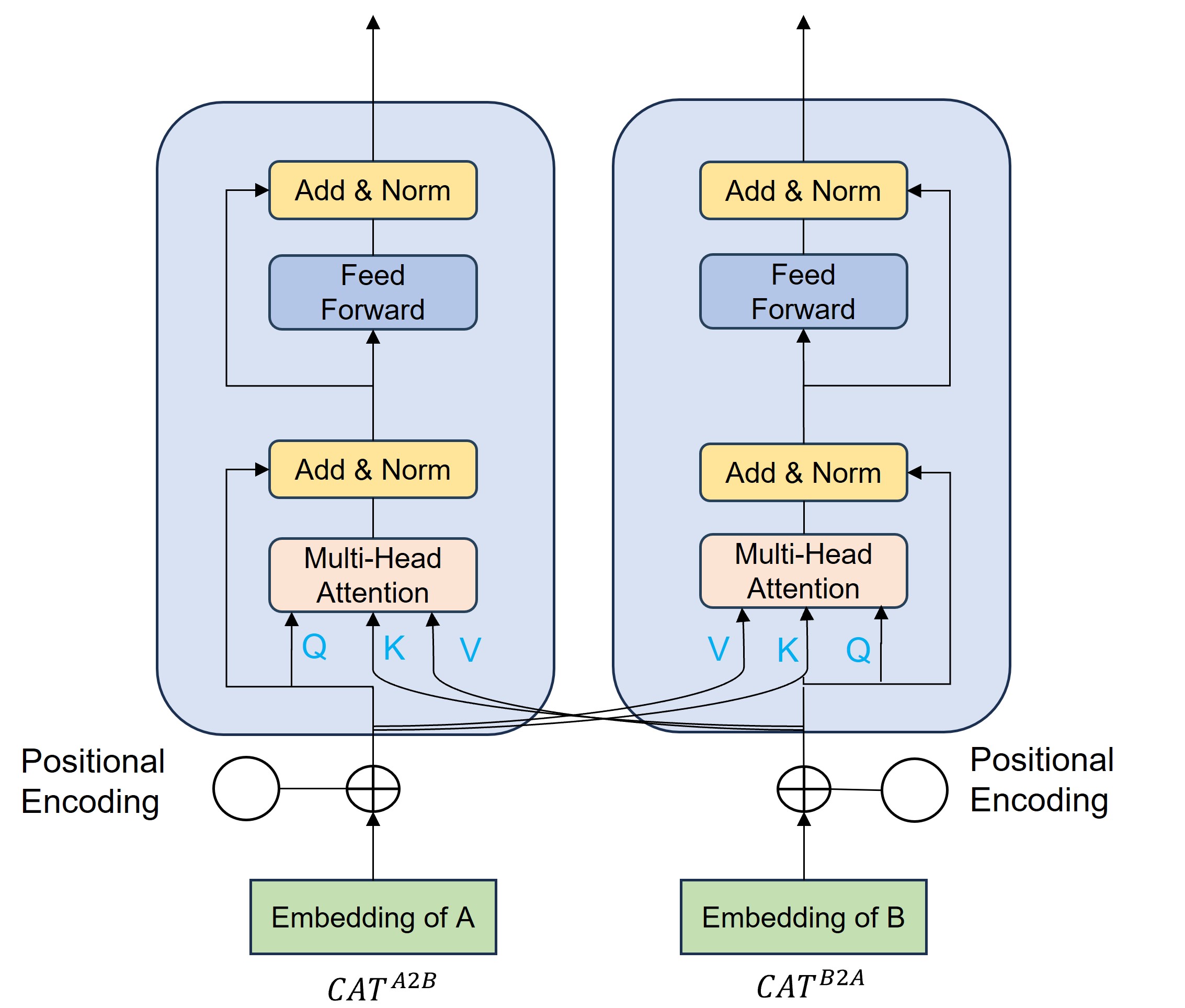}
\caption{ Architecture of Cross-Attention Transformer.}
\label{fig_3}
\end{figure}

A Cross-Attention Transformer (CAT) is an architectural component within transformer models that enables different input sequences to interact and exchange information. 
CAT transfers information from one source to another derived from the concept of self-attention transformers (SAT) \cite{he2023multiple}. As illustrated in Fig.\ref{fig_3}, the CAT takes two input sequences including $A$ and $B$: one sequence serves as the query ($Q$), while the other provides the keys ($K$) and values ($V$). When \(Q^{(A)}\), \(K^{(B)}\) and \(V^{(B)}\) are fed into the transformer, it learns the potential representation from \(R^{(A)}\) to \(R^{(B)}\), denoted \(CAT^{(A2B)}\). In contrast, when \(Q^{(B)}\), \(K^{(A)}\) and \(V^{(A)}\) are inputted into another self-attention transformer, the transformer learns the representation from \(R^{(B)}\) to \(R^{(A)}\), expressed as \(CAT^{(B2A)}\). The general operation of the CAT is formulated as shown in Equation \ref{eq6}.

\begin{equation}
\label{eq6}
CAT(A,B)=CAT(R^{(A)},R^{(B)})
\end{equation}

\begin{figure}[!t]
\centering
\includegraphics[width=9cm,height=11.5cm]{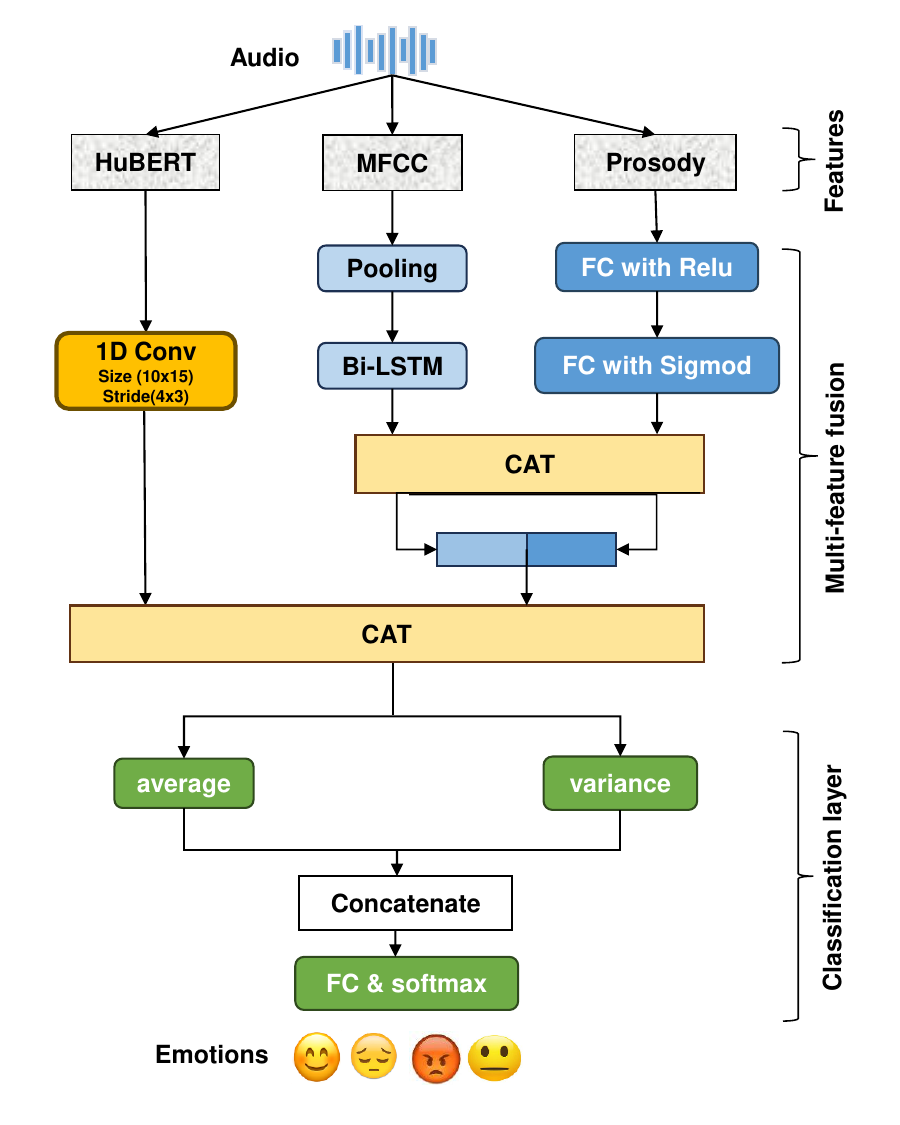}
\caption{Structure of proposed HuMP-CAT.}
\label{fig:overview}
\end{figure}

\HUMP~applies two stages of CAT operation as shown in Fig. \ref{fig:overview}. The detail is illustrated in the following section.

\subsection{Speech Features} 
The SER task relies on extracting global and local speech features. Global features capture statistics properties such as maximum, minimum, standard deviation, and mean, while local features use the temporal dynamics of speech, providing a more detailed understanding of the state of the speaker over time \cite{29}.

\subsubsection{Prosodic Features}
Prosodic information characterizes the intonation and speech rhythm of the speaker \cite{dehak2007modeling}. The most widely used prosodic characteristics are pitch, fundamental frequency, duration, and energy. Prosodic features are more indicative of happiness and anger and less indicative of fear and sadness. Heterogeneous sound features do not affect prosodic features \cite{30}. Research shows \cite{31} that the combination of prosodic characteristics and spectral characteristics has significantly improved SER. 

We adopt the DisVoice \cite{dehak2007modeling,vasquez2018towards} prosodic feature extraction tool to obtain features with a dimension of 103, including the feature of fundamental frequency, energy, and voice/unvoiced duration, and their statistic value such as average, standard deviation, maximum, minimum, skewness, and kurtosis. We suggest referring to this website \footnote{\href{https://github.com/jcvasquezc/DisVoice/tree/master/disvoice/prosody}{https://github.com/jcvasquezc/DisVoice/tree/master/disvoice/prosody}} for details.  

\subsubsection{MFCC Features}
MFCCs are widely used spectral features in speech processing. We refer to the MFCC feature extraction process from \cite{mfcc2023speech}. The raw speech audio (x(t)) is first normalized to minimize noise and disturbances and then divided into 40-ms frames with a frame shift of 20-ms. Each frame is processed using a single Hamming window ($H(k)$) with a frame length (N) of 30 ms.
Next, a discrete Fourier transform is applied to convert the emotional speech signal in the time domain into its frequency domain representation, $Y(k)$, as defined in Equation \ref{eq2}. The power spectrum of the DFT, representing the characteristics of the vocal tract, is calculated using Equation \ref{eq3}. Subsequently, the signal is processed through Mel-frequency triangular filter banks, \(\nabla_m(k) \), as described in Equation \ref{eq4}. Finally, the discrete cosine transform  is applied to the logarithm of the filter bank energy signal to extract $L$ cepstral coefficients, as defined in Equation \ref{eq5}.
\begin{equation}
\label{eq2}
Y(k)=\sum_{t=0}^{N-1}x(t)\times H(k)\times e^{-j2\pi t k/N},0\leq t,k\leq N-1
\end{equation}

\begin{equation}
\label{eq3}
Y_{k}={\frac{1}{N}}|Y(k)|^{2}
\end{equation}

\begin{equation}
\label{eq4}
MT_{m}=\Sigma_{k=0}^{k=1}\nabla_{m}(\mathbf{k})\times Y_{k},m=1,2,...M
\end{equation}

\begin{equation}
\begin{split}
\label{eq5}
M F C C_{i}=\Sigma_{m=1}^{M}\log_{10}(\ MT_{m})\times c o s \left((m+0.5){\frac{i\pi}{m}}\right)\\ for\, i=1,2,\dots L
\end{split}
\end{equation}

Thus, we get 12 MFCC coefficients. As in earlier work \cite{33,34} show that derivative features are important features to characterize speech emotion, we also include 26 first- and second-order derivatives of the MFCC features. In total, we extracted 39 features, including the signal energy. 

The prosodic characteristics (including F0, energy, duration, and their related statistical measures) are passed through 2 fully connected non-linear layers to fit the first input of CAT \(R^{(p)}\). For the MFCC features, we first averaged them and used a Bi-LSTM layer for embedding extraction to obtain the input \(R^{(m)}\). 

In order to improve the generalization performance of \HUMP, the embedding of the prosodic features is combined with the MFCC features through a CAT module,  Next, the second CAT module is used to fuse \(R^{(pm)}\) with the HuBERT embedding \(R^{(h)}\). The two outputs are interacted through two attention layers to obtain the final representation $R$.
\begin{equation}
\label{eq10}
R^{(p m)}=CAT(R^{(p)},R^{(m)})
\end{equation}
\begin{equation}
\label{eq11}
R=CAT(R^{(h)},R^{(pm)})
\end{equation}

The average and variance of the attention output are calculated, which constructs a feature vector of size 64. To classify the embeddings obtained, the AM-Softmax algorithm \cite{zheng2022speakin}is used , which introduces an additional margin to the standard Softmax loss function. The algorithm improves the discriminative power of the model, enabling it to better classify the embeddings of the features into different emotion categories.
\section{Experiments and Results}
\label{sec:experiments}
\subsection{Speech Emotional Datasets}
We evaluate seven emotional speech data sets in five different spoken languages (e.g. English, German, Italian, Spanish, and Chinese). Table \ref{tab:table2} summarizes the datasets. We adopt four categories of emotions including happiness, sadness, neutrality, and anger, as they are the most common across all datasets. The following provides a detailed description of each dataset.
\begin{table*}
\caption{Summary of Speech Emotional Datasets  (M: male, F: Female)\label{tab:table2}}
\centering
\begin{tabular}{ccccc}
\hline
Dataset             & Language      &  Speakers              & Utterance & Emotions   \\ \hline
IEMOCAP\cite{39}    & English       & 10  (5 M \& 5F)         & 5,531 & 9        \\
RAVDESS \cite{41}   & English       & 24 (12 M \& 12 F )     &1,440 & 7  \\
TESS \cite{44}      & English       & 2 (F)                  & 2,800& 7    \\
EMODB\cite{40}      & German        & 10 (5 M \& 5F)         & 535 & 7     \\
 EMOVO \cite{42}    & Italian       & 6  (3 M \& 3 F)         &  588& 7 \\
 MESD\cite{duville2021MESD}      & Spanish       & 11 (2 F, 3 M, and 6 children)  &864 & 5   \\
ESD \cite{ESD2022}  & Chinese       & 10 &420 & 5 \\ \hline
\end{tabular}
\end{table*}
\subsubsection{IEMOCAP}
 The IEMOCAP) \cite{39} is an English multimodal emotional data set which comprises 5,531 utterances from 10 speakers (5 male and 5 female). The actors performed selected emotional scripts and performed nine specific forms of emotions. Most studies have selected improvised data with four emotions (happy, neutral, angry, and sad) and considered excitement as happy emotion to better balance the data. we also relabel excitement samples as happiness.
 
\subsubsection{RAVDESS}
The RAVDESS \cite{41} contains 1440 utterance. The database includes 24 professional actors (12 female, 12 male), who voice two lexically matched statements in North American accent. Speech includes calm, happy, sad, angry, fearful, surprised, and disgusted expressions.

\subsubsection{TESS}
Toronto emotional speech set (TESS) \cite{44} is an English speech database consisting of 2800 samples from two actresses who expressed seven emotions: anger, disgust, fear, happiness, and sadness.

\subsubsection{EMODB}
The Berlin Database of Emotional Speech (EMODB) \cite{40} is a German emotional language dataset with ten actors, including five males and five females. The ten actors simulated seven emotions including the neutral, angry, fearful, happy, sad, disgusted, and bored state. The dataset consists of 535 sentences with 233 sentences of male emotion statements and 302 sentences of female emotion statements.

\subsubsection{EMOVO}
EMOVO \cite{42} is an Italian speech database consisting of 588 records with seven emotions including happiness, sadness, anger, fear, disgust, surprise, and neutral. Three men and three women uttered 14 sentences for each emotion.

\subsubsection{MESD}
The Mexican Emotional Speech Database (MESD) \cite{duville2021MESD} provides single-word utterances in Spanish for anger, disgust, fear, happiness, neutral, and sadness  with Mexican cultural shaping. 864 utterances were uttered by non-professional actors including 3 female, 2 male, and 6 child.

\subsubsection{ESD}
ESD~\cite{ESD2022} is a emotional speech database developed by the National University of Singapore (NUS) and the Singapore University of Technology and Design (SUTD). It consists of 350 parallel utterances spoken by 10 native English speakers in a controlled acoustic environment, covering 5 emotion categories (e.g., neutral, happy, angry, sad, and surprised).

\subsection{Experimental setup}
All audio data sets were resampled at 16 kHz in 16 bits. Each utterance was clipped in the 7-second segments, while we padded a short utterance with its repeated parts. 
We used four common emotions available in all datasets, including happy, neutral, angry, and sad.

The experiments were carried out on an NVIDIA GeForce RTX 4070Ti GPU. To train the models, we chose cross-entropy loss as the loss function and used Adam as the optimizer with an initial learning rate of 1e-3. The models were trained for a maximum of 50 epochs with a batch size of 32. Two evaluation metrics are used for the measurement, including weighted accuracy (WA) and unweighted accuracy (UA).

\subsection{HuBERT Performance of SER on IEMOCAP}
We run the IEMOCAP corpus to compare the SSRL model and speech feature. We use 10-fold cross validation, where 80\% of the speakers (4 males and 4 females) are used to train the model. 10\% of the speakers (one speaker) are used for validation through training, and the remaining 10\% (the other speaker) are used to test the trained model.

\subsubsection{Comparison with other SSRL models}
In order to evaluate the impact of transformer blocks of different speech pre-training models on experimental results, several common SSRL models were used, including HuBERT (HuBERT-base and HuBERT-large), Wav2vec2 (Wav2vec-base and Wav2vec2-large), and WavLM. These models have similar structures, consisting of a CNN encoder and several transformer blocks. These pre-training models were downloaded and called from Huggingface. HuBERT and Wav2vec2 both include base and large versions. The unweighted accuracy results on IEMOCAP are shown in Table \ref{tab:table4}. The result shows that the HuBERT base model has the highest accuracy of 78. 26\%, which outperforms other SSRL models.

\begin{table}[!t]
\caption{Unweighted SER accuracy on IEMOCAP for transformer blocks of different speech pre-trained models\label{tab:table4}}
\centering
\begin{tabular}{ccc}
\hline
Speech pre-trained models & Validation accuracy & Test accuracy \\ \hline
\textbf{HuBERT-base }              & \textbf{81.70  }             & \textbf{78.26 }        \\
HuBERT-large              & 78.53               & 75.25         \\
WavLM                     & 69.36               & 67.80         \\
Wav2vec2-base             & 72.79               & 71.58         \\
Wav2vec2-large            & 70.54               & 68.93         \\ \hline
\end{tabular}
\end{table}

\subsubsection{Effect of speech features}
Next, we run the HuBERT base model on different combination of prosodic features, MFCC, LPCC, and spectrogram features. As shown in Table \ref{tab:table5}, when prosody and MFCC are combined, the experiment has the highest SER accuracy of 74.26\%, which shows that these two features have the bast performance.

\begin{table}[!t]
\caption{Unweighted SER accuracy on IEMOCAP for different combinations of speech features\label{tab:table5}}
\centering
\begin{tabular}{ccc}
\hline
Speech features       & Validation accuracy & Test accuracy \\ \hline
\textbf{Prosody + MFCC}        & \textbf{78.60 }              & \textbf{74.58}         \\
Prosody + LPCC        & 76.01               & 72.45         \\
Prosody + spectrogram & 75.36               & 70.12         \\
MFCC + LPCC           & 68.57               & 65.39         \\
MFCC + spectrogram    & 67.32               & 62.24         \\
LPCC + spectrogram    & 64.82               & 60.09         \\ \hline
\end{tabular}
\end{table}

\subsection{\HUMP~Performance of SER on IEMOCAP }
We compare \HUMP~with other works running on the IEMOCAP data set with four emotions including happiness, sadness, neutrality, and anger. The result is summarized in Table \ref{tab:table3}. 

Li et al.\cite{49} explored the effectiveness of angular softmax loss on
two baselines (CRNN and CNN-TF-GAP models) with different class-agnostic margins. The experiments show the class-specific margin on CNN-TF-GAP baseline performs the best. 

GA-GRU\cite{50} is a graph attention approach in a gated recurrent unit network (GA-GRU), which combines both the long-range attentional time series modeling with the salient frame-wise graph structure within an emotional utterance, and achieves the accuracy of 63.8\% WA and 62.27\% UA.

HFGM\cite{51}  is a hierarchical grained and feature mode, is a hierarchical grained feature, which uses RNN to process the frame-level and utterance-level structures of acoustic samples so that it can capture features of different granularities and improve the
sensitivity of the model. 

HuBERT-LinearLayer\cite{48} first applied HuBERT to get representation and then tested on three DNN-based classifiers: Pooling + Linear Layer, CNN with Self-Attention, and CT-Transformer, respectively. The result shows that the Pooling + Linear Layer has a better accuracy of 65.60\%.

TIM-Net\cite{46} extracted the MFCC features by applying the Hamming window to each
speech signal with a frame length of 50 ms and a 12.5 ms shift. The 39 extracted features were fed into Bi-Temporal-Aware Blocks (TABs). The 10-fold cross-validation with 90\% and 10\% samples in train and test sets was applied.

AMSNet\cite{45}  applied a spatial CNN with the squeeze-and-excitation block (SCNN) to extract spectrogram representations and an Attention Bi-LSTM  to extract temporal features. The training and validation set of the IEMOCAP was split into 80\% and 20\% of the total samples. The accuracy is 69.22\% (WA) and 70.51 \% (UA). 

Wav2Vec2 P-TAPT\cite{47} is a Wav2Vec2-based fine-tuning approach using adaptive pseudo-label task pretraining (P-TAPT), which achieved a significant improvement of more than 5\%  in unweighted accuracy (UA) over state-of-the-art performance on IEMOCAP.

\begin{figure}[!t]
\centering
\includegraphics[width=1\linewidth]{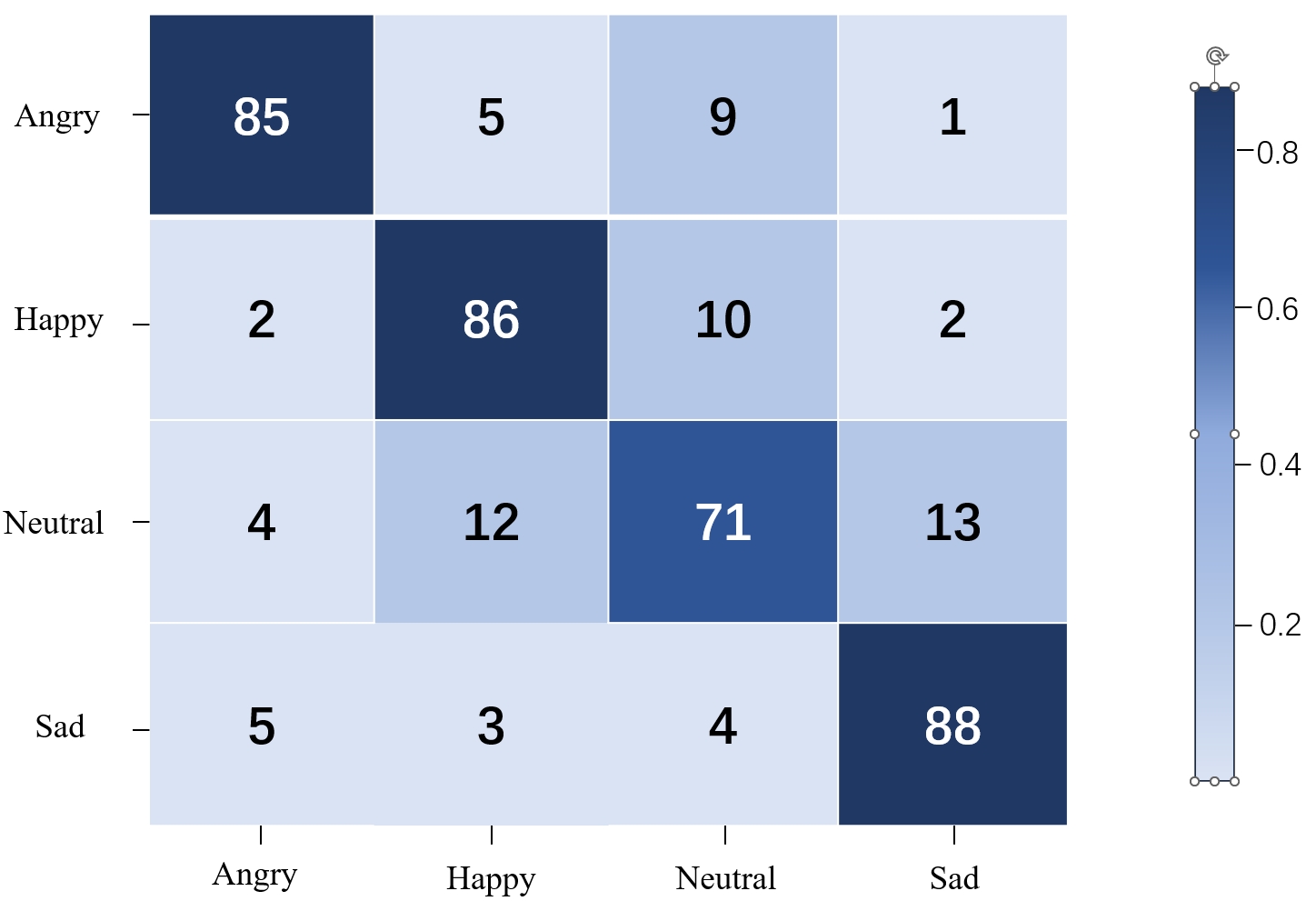}
\caption{Confusion matrix of \HUMP~on IEMOCAP corpus.}
\label{fig_6}
\end{figure}

\HUMP~achieved the accuracy of the UA of 82.45\% and the accuracy of the WA of 81.27\%, which outperforms other state-of-the-art learning-based feature fusion and transfer methods as shown in Table \ref{tab:table3}. Fig.\ref{fig_6} shows the confusion matrix, \HUMP~has a higher SER accuracy above 80\% for strong emotions (e.g., angry, happy, and sad) except for the neutral emotion with a lower accuracy of 71\%. we use this trained model for the following evaluation and comparison. 

\begin{table}[!t]
\caption{Comparison of the proposed method with other works on IEMOCAP corpus}
 \label{tab:table3} 
\centering
\begin{tabular}{ccccc}
\hline
Method (year)                                     & UA(\%) & WA(\%) &  \\ \hline
\textbf{HuMP-CAT }                           & \textbf{82.45}  & \textbf{81.27}  &  \\ 
CNN-improved + Angular Softmax (2019)  \cite{49}  & 64.80  & 73.33  &  \\
GA-GRU (2020) \cite{50}                           & 62.27  & 63.80  &  \\
HFGM (2020)\cite{51}                            & 66.54  & 70.48  &  \\ 
HuBERT-LinearLayer (2022) \cite{48}              & 65.60  & \_     &  \\
TIM-Net (2023) \cite{46}                          & 69  & 68.29  &  \\
AMSNet (2023) \cite{45}                           & 70.51  & 69.22  &  \\
Wav2Vec2 P-TAPT (2023) \cite{47}               & 74.3   & \_     &  \\
\hline

\end{tabular}
\end{table}

\subsection{\HUMP~Performance for CLSER}
We adopt a transfer learning approach, fine-tuning the model pre-trained on the IEMOCAP dataset using a small subset of the target datasets. Specifically, we evaluated our model on seven speech emotion datasets: EMODB, MESD, EMOVO, RAVDESS, SAVEE, TESS, and ESD. For fine-tuning, we used 20\% of the speakers' speech from EMODB, one third of the speakers' speech from EMOVO, and 10\% of the data from the remaining datasets. We compare with other state-of-the-art work in the table\ref{tab:table6}.

Latif et al.\cite{52} applied a generative adversarial network (GAN)-based model to encode the given data into the underlying feature structure and then uses SVM for emotion classification.  By including the EMODB, URDU, SAVEE, and EMOVO dataset, the study adopted the one-language data-out approach, and the remaining three corpora are mixed and used to train the model. Table\ref{tab:table6} lists the best performance of the GAN-SVM approach. 
Meng et al.\cite{54} introduced an architecture ADRNN which applied dilated CNN with residual block and BiLSTM based on the attention mechanism. ADRNN uses the log-mel spectrogram as the feature. The experiment adopted IEMOCAP as a training set and tested the model in the EMODB database, showing an accuracy of 63.84\%.
The VACNN + BOVW approach \cite{53} designed a visual attention convolutional neural network
(VACNN) pre-trained with TESS and RAVDESS datasets by extracting the feature on a log-mel spectrogram using a bag of visual words(BOVW). The pretrained model is fine-tuned with the target dataset through five-fold cross validation.
 \cite{57} compared the performance of an ensemble learning approach against traditional machine learning including RF, SVM, and Decision Tree on four corpora (e.g., SAVEE, URDU, EMO-DB, and EMOVO). This study tested a classifier performance trained in one corpus with data from another corpus to evaluate its efficiency for CLSER and showed that ensemble learning performs better. 
  All of the works used either low-level acoustic features or log-mel spectrogram. 
 
 We also compared \HUMP with other SSRL approaches. Zehra et al.
 Ahn et al. \cite{56} introduced Few-shot Learning and Unsupervised Domain Adaptation (FLUDA), which trains an embedding and a metric module to project utterances into a meaningful shared feature space and discern class differences, respectively. During training, an auxiliary module is incorporated to differentiate between real and pseudo-labeled samples. The proposed method uses IEMOCAP and CREMA-D \cite{cao2014crema} as source corpora. In experiments with EMODB as the target corpus, FLUDA achieved an accuracy of 56.8\%.
MDAT\cite{zaidi2024enhancing} is a multimodal model that leverages pre-trained models including RoBERTa \cite{liu2019bert} with wav2vec2-XLS-R \cite{babu2021xls} to extract multi-language speech and text embedding
EmoBox \cite{ma2024emobox}  provided toolkit and benchmark for Multilingual Multi-corpus SER. The cross-corpus SER results on 4 datasets (e.g., IEMOCAP, MELD (English),
RAVDESS and SAVEE) with fully balanced test sets. The best result with the Whisper large v3 \cite{radford2023robust} demonstrates the best performance among other SSRL such as HuBERT base/large, WavLM base/large, data2vec base/large, and data2vec 2.0 base/large.

Table \ref{tab:table6} lists the performance of our proposed model \HUMP, which achieves an impressive average accuracy of 78.75\%  across the seven diverse data sets. 
Notably, the ESD test (Chinese) showed a lower accuracy of 60.35\%, likely due to significant linguistic and acoustic differences between English and Chinese, but in the worst case,  \HUMP~still perform better than others. 

\begin{figure}[!t]
\centering
\includegraphics[width=7.5cm,height=6cm]{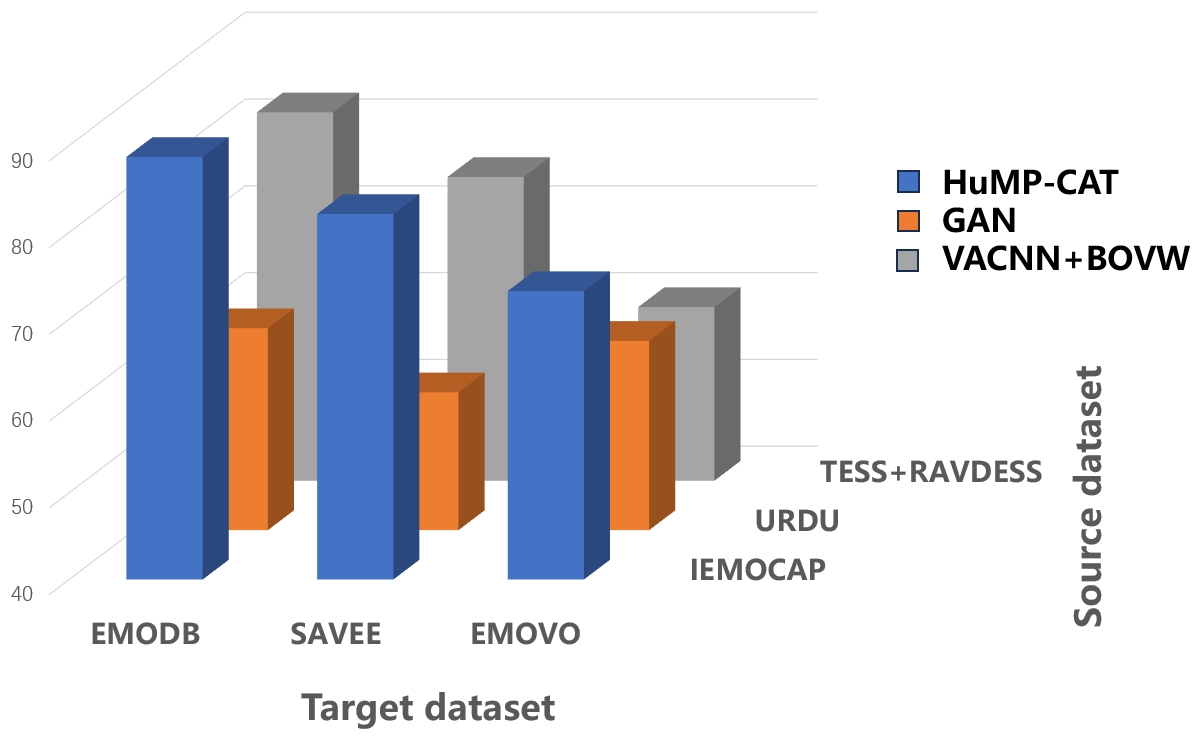}
\caption{Comparison of three methods on the same target dataset.} 
\label{fig_7}
\end{figure}

Fig.\ref{fig_7} illustrates the comparison of accuracy between \HUMP~and two other representative methods GAN \cite{52} and VACNN+BOVW \cite{53}, on the same target data set. Although the source datasets used by these methods differ, their results provide useful reference points. As shown, \HUMP~significantly outperforms GAN and achieves slightly better results than VACNN+BOVW across the three target datasets, highlighting the effectiveness of our proposed approach.

\begin{table*}[!t]
\caption{Comparison With the state-of-the-arts CLSER}
\label{tab:table6}
\centering
\begin{tabular}{ccccc}
\hline
Method (year)    & Source dataset   & Features   & Target Dataset   & UA                                                          \\ \hline
\multirow{7}{*}{\textbf{HuMP-CAT}} & \multirow{7}{*}{\textbf{IEMOCAP}} & \multirow{7}{*}{ }&EMODB  &88.69  \\
                          &              &    Speech representations        & RAVDESS  & 83.56    \\
                          &             &     MFCC        & TESS     & 85.45       \\
                          &             &    Low-level acoustic features  &  SAVEE   & 77.91      \\
                          &             &             &   EMOVO      & 79.48                      \\
                          &             &             & MESD     & 72.84            \\
                          &              &            & ESD      & 60.35        \\ \hline
                          
                          &\multirow{3}{*}{EMODB, URDU, SAVEE, or URDU}   &\multirow{3}{*}{ }  & EMODB  & 63.25\\
GAN-SVM (2019) \cite{52}    &  & Low-level acoustic features  & SAVEE & 55.12\\
                            &  &Latent code  & EMOVO & 61.8 \\ \hline
ADRNN (2019) \cite{54}     & IEMOCAP      & Log-mel spectrogram   & EMODB   & 63.84  \\\hline
                          &\multirow{3}{*}{TESS+RAVDESS}  &\multirow{3}{*}{}  & EMODB & 86.92  \\
VACNN+BOVW (2020) \cite{53} & &  Log-mel spectrogram & SAVEE & 75.00 \\
                           &   &   & RAVDESS  &  83.33   \\ \hline
Ensemble (2021) \cite{57}       & EMOVO   &  MFCC, Low-level acoustic features   & URDU   & 62.5   \\\hline
FLUDA (2021) \cite{56}            & CREMA-D    & Speech representations  & EMODB    & 56.8  \\\hline
\multirow{2}{*}{MDAT (2024)\cite{zaidi2024enhancing} }  &\multirow{2}{*}{IEMOCAP}   & Speech representations  &   EMOVO   & 85.51\\
    &   & Text embedding & EMODB  & 42.48\\\hline  

\multirow{3}{*}{EmoBox (2024)\cite{ma2024emobox} }  &\multirow{3}{*}{IEMOCAP}   & \multirow{3}{*}{Speech representations}  &   RAVDESS    & 48.12\\
    &   &  & SAVEE & 49.30\\
     &   &  &  MELD & 51.42\\\hline  
\end{tabular}
\end{table*}

\subsection{Impact of Speech features with Different Dataset on CLSER}

To assess the effectiveness of \HUMP~in incorporating prosodic and MFCC features in CLSER, we compared the performance of the transfer learning method with and without these features, while consistently including the HuBERT model. The training and testing configurations are consistent with those described in the previous section. As summarized in Table \ref{tab:table7}, the results indicate that \HUMP~with both prosodic and MFCC features consistently outperforms the model with one of the two features alone.  

\begin{table}[!t]
\caption{Accuracy of \HUMP~with different speech features on test data\label{tab:table7}}
\centering
\begin{tabular}{c|ccc}
\hline
Target Dataset  & MFCC   & Prosody   & MFCC \& Prosody     \\ \hline
EMODB  & 82.54    & 83.36   & 88.69     \\
EMOVO  & 70.23    & 68.94   & 75.48     \\
MESD   & 66.21    & 65.30   & 72.84     \\
RAVDESS  & 81.95  & 83.12   & 83.56     \\
TESS     & 76.22  & 78.80   & 85.45      \\ \hline
\end{tabular}
\end{table}

\section{Conclusion}
\label{sec:concusion}
In this paper, we proposed a novel CLSER framework, \HUMP, designed for low-resource emotional speech datasets. The model leverages convolutional layers and cross-attention mechanisms to effectively fuse three key features: outputs from the HuBERT transformer block, MFCCs, and prosodic features. This fusion enhances the model's adaptability to diverse language datasets. \HUMP~was pre-trained on the IEMOCAP corpus as the source domain and fine-tuned on seven target datasets—EMODB, EMOVO, MESD, TESS, SAVEE, RAVDESS, and ESD—spanning five spoken languages: German, English, Spanish, Chinese, and Italian.

To evaluate its performance on low-resource speech emotion datasets, we fine-tuned the model using only a small subset of each target dataset. \HUMP~achieved an impressive average accuracy of 78.75\% across the seven datasets, with notable results such as 88.69\% on EMODB and 79.48\% on EMOVO. Our extensive evaluation demonstrates that \HUMP~outperforms existing methods across multiple target languages, demonstrating its robustness and effectiveness.

For future work, we aim to expand the source training dataset by including additional emotional speech corpora to further improve the generalization and cross-linguistic adaptability of the model.

\bibliographystyle{IEEEtran}
\bibliography{main}

\end{document}